\documentclass[journal,10pt]{IEEEtran}
\usepackage{float}
\usepackage{xcolor,soul,framed} 
\usepackage{mathrsfs}
\usepackage{amsfonts}
\usepackage{bbm}
\usepackage{amssymb}
\usepackage{mwe,subcaption,tikz}
\usepackage{array}
\usepackage{mdwmath}
\usepackage{mdwtab}
\usepackage{eqparbox}
\usepackage{url}
\usepackage{cite}
\usepackage{amsmath}
\usepackage[noend]{algpseudocode}
\usepackage{algorithmicx,algorithm}
\usepackage{caption}
\usepackage[utf8]{inputenc}

\usepackage{mathtools}
\DeclareMathOperator{\Tr}{Tr}
\DeclareMathOperator{\Diag}{Diag}

\tikzset{boximg/.style={remember picture,red,thick,draw,inner sep=0pt,outer sep=0pt}}
\colorlet{shadecolor}{yellow}

\begin{document}
\title{Decentralized Semantic Communication and Cooperative Tracking Control for a UAV Swarm over Wireless MIMO Fading Channels }%
\author{Minjie~Tang, \IEEEmembership{Member, IEEE},  Chenyuan Feng$^*$, \IEEEmembership{Member, IEEE}, and Tony Q. S. Quek, \IEEEmembership{Fellow, IEEE} 
\thanks{* denotes the corresponding author.}
\thanks{Minjie Tang and Chenyuan Feng are with the Department of Communication Systems,  EURECOM, France (e-mail: Minjie.Tang@eurecom.fr; Chenyuan.Feng@eurecom.fr). }
\thanks{Tony Q. S. Quek is with  the Information Systems Technology and Design Pillar, Singapore University of Technology and Design, Singapore (e-mail:  tonyquek@sutd.edu.sg) }
\thanks{This work has been submitted to the IEEE for possible publication. Copyright may be transferred without notice, after which this version may no longer be accessible.}
}

\maketitle

\begin{abstract}
This paper investigates the semantic communication and cooperative tracking control for an  UAV swarm comprising a leader  UAV and a group of follower UAVs, all interconnected via unreliable wireless multiple-input-multiple-output (MIMO) channels. Initially, we develop a dynamic model for the UAV swarm that accounts for both the internal interactions among the cooperative follower UAVs  and the imperfections inherent in the MIMO channels that interlink the leader and follower UAVs. Building on this model, we incorporate the power costs of the UAVs and formulate  the communication and cooperative tracking control challenge as a drift-plus-penalty optimization problem. We then derive a closed-form optimal solution that maintains a decentralized semantic architecture, dynamically adjusting to the tracking error costs and local channel conditions within the swarm. Employing Lyapunov drift analysis, we establish closed-form sufficient conditions for the stabilization of the UAV swarm's tracking performance.   Numerical
results demonstrate the significant enhancements in our proposed scheme over  various state-of-the-art methods.
\end{abstract}


\section{Introduction}
\begin{figure}
    \centering
    \includegraphics[height=4.4cm,width=8cm]{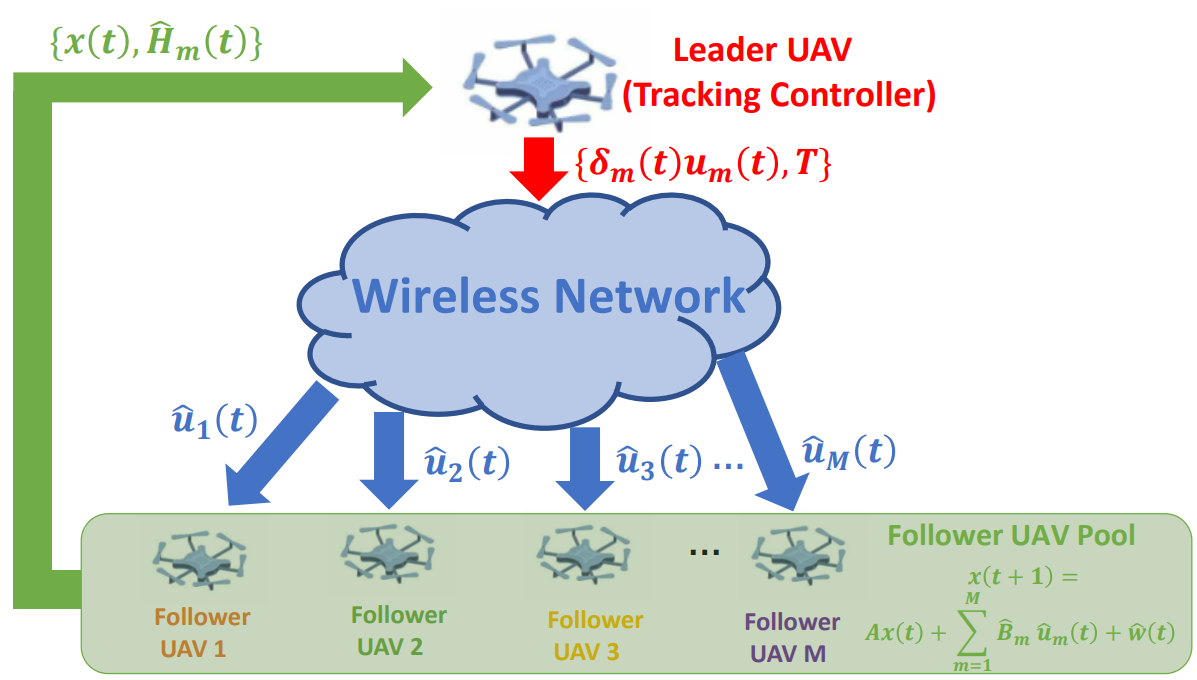}
    \caption{Typical architecture of a UAV swarm over the wireless network.}
    \label{fig1}
\end{figure}

\IEEEPARstart{C}ooperative tracking control for  unmanned aerial vehicle (UAVs) swarms has garnered substantial interest across both the industrial and academic realms, owing to its broad applications in fields such as surveillance and agricultural monitoring \cite{liu2022resource,khalid2023control}.
A typical UAV swarm, comprising a group of \emph{follower UAV agents}  and a \emph{leader controller UAV}, is depicted in Fig. \ref{fig1}. The leader UAV monitors the real-time states of the follower UAVs—including their position, speed, and angular velocity—and intermittently generates tracking control signals that are transmitted to the follower UAVs via an unreliable wireless network. Upon reception of these control directives, the follower UAVs adjust their states to conform to predetermined target profiles.
The wireless network connecting the follower UAVs and the leader  UAV is susceptible to a myriad of impairments, such as signal  fading and channel noise, which has the potential to markedly degrade the tracking control efficacy of the UAV swarm.

Tracking controller design for UAV swarms over wireless network entails considerable challenges. Firstly, most existing research on UAV control assumes static communication channels within UAV swarms. For instance, UAV tracking control laws that leverage the pole placement technique and the proportional-integral-derivative (PID)-based approach have been developed in \cite{tran2022angular} and \cite{su2024fixed, bertran2016tradeoff, lv2021adaptive}, respectively. However, these methods are heuristic and lack optimality. To mitigate reliance on trial-and-error methods, a linear optimal tracking control law utilizing Riccati equations to derive optimal control gains for UAV swarms was introduced in \cite{cohen2020finite, luo2024synchronous}. It is important to note that all the aforementioned works assume static channels in UAV swarms.  Brute-force applications of these control laws to  time-varying  fading channels can severely impair tracking performance. While several studies, such as \cite{singh2022path}, consider random wireless channels in UAV swarms, they oversimplify the wireless network to packet dropout channels with finite channel state information (CSI) realizations. Extending these models to accommodate more general wireless fading channels remains a significant challenge.

Secondly, many existing works on controller  design for UAV swarms  overlooked the power costs at UAVs, focusing primarily on tracking stability performance. To address power efficiency, recent studies have integrated communication and tracking controller design for UAV swarms. For instance, in \cite{huang2021stochastic}, a periodic communication and control policy was considered, where UAVs communicate  updated tracking control signals at regular intervals. However, such an approach fail to account for the real-time states of UAVs.  When the discrepancy between the target and real-time UAV states are significant, a long activation period can lead to tracking instability. Conversely, short activation periods, when the differences are minimal, may result in unnecessary power consumption. Recent advancements in semantic communications, as explored in \cite{luo2022semantic,pokhrel2022understand,su2023semantic,GSS2023}, focus on transmitting crucial application-specific information. In the context of tracking control for UAV swarms, this involves prioritizing control commands that significantly influence tracking stability. For instance, state-dependent semantic control, as proposed in \cite{wei2020dynamic,zhang2022distributed,wu2023qos,liu2024decentralized}, triggers transmissions among UAVs based on significant tracking errors. However, in scenarios involving wireless channels, it is critical to also embrace the channel state in both communication and controller design. Consequently, semantic communication for UAV swarms  over complex wireless communication environments requires further investigation.

Motivated by these issues, we investigate the decentralized semantic communication and cooperative tracking control for  UAV swarms over the wireless MIMO fading channels. The key contributions are summarized as follows.
\begin{itemize}
\item \textbf{Semantic Communication for UAV Swarms over MIMO Fading Channels.} We introduce an efficient  communication strategy by taking aware both the power cost and tracking stability of UAV swarms. The solution has an semantic structure that adapts to the real-time tracking error costs and the  local channel conditions.

\item \textbf{Semantic Cooperative  Tracking Control for UAV Swarms over MIMO Fading Channels.}  To effectively control the UAV swarm under random wireless network, the tracking control strategy for UAVs must adapt to the UAV state and the instantaneous CSI. Traditional approaches involve solving complex systems of coupled Bellman equations in an augmented state space \cite{tang2022online}. Thus, we introduce a low-complexity control algorithm that minimizes Lyapunov drift across the multi-agent UAV swarm under general wireless MIMO fading channels. Our control policy is semantically structured, relying solely on the UAV states and local CSIs in the swarm.

\item \textbf{Closed-form Tracking Control Performance Analysis.} Analyzing the tracking stability performance of UAV swarms is challenging due to state-dependent communication and control solutions, as well as the time-varying nature of the wireless channels among the UAVs. In this article, we utilize the Lyapunov analysis method to establish a closed-form sufficient condition for tracking stability in UAV swarms operating over MIMO fading channels.
\end{itemize}

\section{System Model}
\subsection{Dynamic Model}
A typical UAV swarm comprises $M\in\mathbb{Z}_+$ geographically distributed follower UAVs and a leader UAV, interconnected through  an unreliable wireless network, as shown in Fig. 1.  We assume that the leader UAV and the follower UAVs are equipped with $N_t$ transmission antennas and  $N_r$ receiving antennas, respectively. The physical  process for each $m$-th follower UAV is described by a set of first-order coupled equations as follows:
\begin{equation}\label{eq1}
    \mathbf{x}_m(t+1)=\mathbf{A}_{mm}\mathbf{x}(t)+\mathbf{B}_m\widehat{\mathbf{u}}_m(t)+\mathbf{w}(t), m\in\left\{1, 2,...,M\right\},
\end{equation}
where $\mathbf{x}_m(t)=[p_{m,x}(t),p_{m,y}(t), p_{m,z}(t), v_{m,x}(t),v_{m,y}(t),$ $v_{m,z}(t),$ $a_{m,x}(t),a_{m,y}(t),a_{m,z}(t)]^T\in\mathbb{R}^{9\times 1}$ is the state of the $m$-th follower UAV. $p_{m,n}(t)\in\mathbb{R}, v_{m,n}(t)\in\mathbb{R}$ and $a_{m,n}(t)\in\mathbb{R}$ are the position, speed, and angular speed of $m$-th UAV at positive $n$-th direction, respectively. $\mathbf{A}_{mm}\in\mathbb{R}^{9\times 9}$ and $\mathbf{B}_m\in\mathbb{R}^{9\times N_{r}}$  are the internal transition matrix and actuation matrix for $m$-th follower UAV, respectively. $\widehat{\mathbf{u}}_m(t)\in\mathbb{R}^{N_{r}\times 1}$ is the received tracking control signal at $m$-th follower UAV. $\mathbf{w}_m(t)\sim\mathcal{N}(0, \mathbf{W}_m)$ is the additive plant  noise at $m$-th follower UAV with finite covariance matrix $\mathbf{W}_m\in\mathbb{S}^{9}$.

By aggregation, the UAV swarm follows the dynamics as follows:
\begin{equation}\label{eq2}
    \mathbf{x}(t+1)=\mathbf{A}\mathbf{x}(t)+\sum_{m=1}^M\widehat{\mathbf{B}}_m\widehat{\mathbf{u}}_m(t)+\widehat{\mathbf{w}}(t),
\end{equation}
where $\mathbf{x}(t)=[\mathbf{x}_1^T(t),...,\mathbf{x}_M^T(t)]^T\in\mathbb{R}^{9M\times 1}$ is the global state for the UAV swarm. $\mathbf{A}=\tiny\begin{bmatrix}
    \mathbf{A}_{11}&\cdots&\mathbf{A}_{1M}\\\vdots &\vdots &\vdots\\ \mathbf{A}_{M1} &\cdots &\mathbf{A}_{MM}
\end{bmatrix}\in\mathbb{R}^{9M\times 9M}$ is the global transition matrix, where $\mathbf{A}_{m,n}$ characterizes the dynamic relationship between $m$-th and $n$-th follower UAVs. $\hat{\mathbf{B}}_m=[\mathbf{0}_{N_{r}\times (9m-9)},\mathbf{B}_m^T,\mathbf{0}_{N_{r}\times (9M-9m)}]^T\in\mathbb{R}^{9M\times N_{r}}$ is the $m$-th global actuation matrix\footnote{Throughout the paper, 
$\mathbf{0}_a$, $\mathbf{0}_{a\times b}$, $\mathbf{I}_a$, and $\mathbf{I}_{a\times b}$ denote an $a\times a$ matrix where all elements are zero, an $a\times b$  matrix with all zero elements, an $a\times a$  identity matrix and an $a\times b$  matrix where all elements are one, respectively.}. $\widehat{\mathbf{w}}(t)\sim\mathcal{N}(\mathbf{0}_{9M\times 1},\Diag(\mathbf{W}_1,...,\mathbf{W}_M))$ is the global additive plant noise.

\subsection{Wireless Communication Model}
The leader UAV observes the states $\mathbf{x}(t)$ of the follower UAVs via the depth-of-field (DOF) camera\cite{gallego2017event}, and generates  remote tracking control signal $\mathbf{u}_m(t)\in\mathbb{R}^{N_{t}\times 1}$ for each $m$-th follower UAV. The signal $\mathbf{u}_m(t)\in\mathbb{R}^{N_{t}\times 1}$ will be conveyed to the $m$-th follower UAV over wireless MIMO fading channels. At each $m$-th follower UAV, 
the received signal $\widehat{\mathbf{u}}_m(t)\in\mathbb{R}^{N_{r}\times 1}$  is given by:
\begin{equation}\label{eq3}
    \begin{split}
        \widehat{\mathbf{u}}_m(t)=\delta_m(t)\mathbf{H}_{m}(t)\mathbf{u}_m(t)+\mathbf{v}_m(t), 1\leq m\leq M,
    \end{split}
\end{equation}
where 
$\delta_{m}(t)\in\left\{0, 1\right\}$ is the communication variable that indicates the communication activity between the leader UAV and $m$-th follower UAV. $\mathbf{v}_m(t)\sim\mathcal{N}(\mathbf{0}_{N_r\times 1},\mathbf{1}_{N_{r}})$ is the additive channel noise at $m$-th follower UAV. $\mathbf{H}_m(t)\in\mathbb{R}^{N_r\times N_t}$
is the wireless MIMO channel fading between the leader UAV and the $m$-th follower UAV. It remains constant within each timeslot and is i.i.d. over follower UAVs and timeslots. Each element of $\mathbf{H}_m(t)$ follows a Gaussian distribution with zero mean and unit variance.

\subsection{Performance Metric}
Let the target state $\mathbf{r}(t)\in\mathbb{R}^{9M\times 1}$ evolve according to 
\begin{equation}\label{eq4}
    \mathbf{r}(t+1)=\mathbf{G}\mathbf{r}(t),
\end{equation}
where $\mathbf{G}\in\mathbb{R}^{9M\times 9M}$ is the target  transition matrix.
The primary objective for the leader UAV  is to drive the state $\mathbf{x}(t)$ to track the target profile $\mathbf{r}(t)\in\mathbb{R}^{9M\times 1}$ by designing
the control signals $\left\{\mathbf{u}_m(t)\right\}$. Specifically, we have the following definition on the tracking stability of the UAV  swarm.

\emph{Definition 1: (Tracking Stability of the UAV Swarm)} The UAV swarm is tracking stable if
\begin{equation}\label{eq5}
    \!\!\!\!\!\limsup_{T\rightarrow\infty}\frac{1}{T}\sum_{t=1}^T\mathbb{E}_{\mathbf{w}(t),\left\{\mathbf{v}_m(t),\delta_m(t),\mathbf{H}_m(t)\right\}}\left\{\|\mathbf{x}(t)-\mathbf{r}(t)\|^2\right\}<\infty.
\end{equation}

In the subsequent sections, we leverage the Lyapunov optimization technique to derive a novel decentralized semantic communication and control solution that achieves  tracking stability of the UAV  swarm.

\section{Problem Formulation and Proposed Method}

\subsection{Decentralized Lyapunov Optimization Formulation}
We adopt Lyapunov theory to obtain the control solutions as well as communication policies for the UAV swarm. Since our target is to maintain the tracking stability of the UAV swarm, we  define a Lyapunov function with respect to (w.r.t.) the tracking error cost $\Sigma(t)=(\mathbf{x}(t)-\mathbf{r}(t)(\mathbf{x}(t)-\mathbf{r}(t)^T$, as follows:
\begin{equation}\label{eq6}
L(\Sigma(t))=\Tr(\Sigma(t)).
\end{equation}

The associated Lyapunov drift  for $L(\Sigma(t))$  is defined to be:
\begin{equation} \label{eq7}
\!\!\!\!\Gamma(\Sigma(t))=\mathbb{E}_{\mathbf{w}(t),\left\{\mathbf{v}_m(t),\mathbf{H}_m(t),\delta_m(t)\right\}}[L(\Sigma(t+1))-L(\Sigma(t))|\Sigma(t)].
\end{equation}

To analyze the Lyapunov drift \eqref{eq7}, we revisit the dynamics for states $\mathbf{x}(t)$ and $\mathbf{r}(t)$ in Section II. Specifically, we
substitute \eqref{eq2}, \eqref{eq3}, \eqref{eq4}, and \eqref{eq6} into \eqref{eq7}
and consider the state-feedback control law $ \mathbf{u}_m(t)=-\mathbf{K}_m(t)(\mathbf{x}(t)-\mathbf{r}(t))$ for follower UAVs, where $\mathbf{K}_m(t)\in\mathbb{R}^{N_t\times 9M}$.
This leads us to the following theorem on the Lyapunov drift.

\newtheorem{theorem}{Theorem}
\begin{theorem}
\emph{(Lyapunov Drift)}  Let the singular value decomposition (SVD) of $\mathbf{A}$ and $\mathbf{G}$ be $\mathbf{A}=\mathbf{U}_1\Pi_1\mathbf{V}_1^T$ and $\mathbf{G}=\mathbf{U}_2\Pi_2\mathbf{V}_2^T$, respectively, where $\mathbf{U}_i\in\mathbb{R}^{9M\times 9M}$ and $\mathbf{V}_i\in\mathbb{R}^{9M\times 9M}$ are unitary matrices, and
$\Pi_i=\Diag(\pi_{i,1},...,\pi_{i,9M})\in\mathbb{S}^{9M}$. Define  $\pi_m=\pi_{2,m}\mathbf{1}_{|\pi_{1,m}|>|\pi_{2,m}|}+\pi_{1,m}\mathbf{1}_{|\pi_{2,m}|>|\pi_{1,m}|}$, where $\mathbf{1}_{\left\{\mathcal{A}\right\}}\in\left\{0, 1\right\}$ is an indicator function that equals 1 if and only if the event $\mathcal{A}$ holds true.  Let
$\Pi=\Diag(\pi_1,...,\pi_{9M})\in\mathbb{S}^{9M}$ and $\alpha=2\max\left\{\|\mathbf{A}\|^2,\|\mathbf{G}\|^2\right\}$.
The Lyapunov drift as defined in \eqref{eq7} can be expressed as follows:
\begin{equation}\label{eq8}
\begin{split}
    &\!\Gamma(\Sigma(t))\leq \Tr(\mathbf{W})-\Tr(\Sigma(t))+\alpha\Tr(\Sigma(t))-\mathbb{E}[2\sum_{m=1}^M\Tr(\\&\delta_m(t)\mathbf{B}_m\mathbf{H}_m(t)\mathbf{K}_m(t)\Sigma(t)\Pi)+\sum_{m=1}^M M\delta_m(t)\Tr(\Sigma(t)(\mathbf{B}_m\\&\mathbf{H}_m(t)\mathbf{K}_m(t))^T(\mathbf{B}_m\mathbf{H}_m(t)\mathbf{K}_m(t)))|\Sigma(t)]+\sum_{m=1}^M\Tr(\mathbf{B}_m\mathbf{B}_m^T).
\end{split}
\end{equation}
\end{theorem}

\begin{IEEEproof}
See Appendix A.
\end{IEEEproof}

Note that the negative Lyapunov drift stabilizes the plant state by pulling it back to the equilibrium point. Therefore, it is beneficial to optimize the tracking control variable 
 $\left\{\mathbf{K}_{m}(t)\right\}$ and communication variable $\left\{\delta_{m}(t)\right\}$ by minimizing the Lyapunov drift \eqref{eq8}. Since these variables are decoupled in \eqref{eq8},  we can optimize $\mathbf{K}_m$ and  $\delta_m(t)$ independently for each $m$-th follower UAV. Further
considering the communication consumption in the UAV swarm, the dynamic control and communication problem for each UAV can be formulated as follows.

\emph{Problem 1: (Decentralized  Lyapunov Optimization for the UAV Swarm})
For a given $\Sigma(t)$ and $\mathbf{H}_m(t)$, the  communication solution $\delta_m^*(t)$ and the control solution $\mathbf{K}_m^*(t)$ for the $m$-th follower UAV can be obtained by solving the problem below.
\begin{equation}\label{eq9}
\begin{split}
\min_{\delta_m(t),\mathbf{K}_m(t)}& -2\delta_m(t)\Tr(\mathbf{H}_m(t)\mathbf{K}_m(t)\Sigma(t)\Pi)+M\delta_m(t)\\&\Tr(\Sigma(t)(\mathbf{H}_m(t)\mathbf{K}_m(t))^T(\mathbf{H}_m(t)\mathbf{K}_m(t)))+\delta_m(t)\\&(P_{on}+\gamma\Tr(\mathbf{K}_m(t)\mathbf{K}^T_m(t)))
\\& \!\!\!\!\!\!\!\!\!\! \!\!\!\!\!\!\!\!\!\!\text{s.t.}\,\,\,\,\,\,\,\,\,\,\,\,\,\,\delta_m(t)\in\left\{0, 1\right\}.
\end{split}
\end{equation}
where $\gamma\geq 0$ is the communication  price among the UAVs and $P_{on}\geq 0$ is the activation power consumption for UAVs.

Although Problem 1 is a convex optimization problem, it is still challenging to obtain a closed-form solution due to the integer constraints.

\subsection{Decentralized Semantic Solution}
Consider the SVD of  $\delta_m(t)\mathbf{B}_m\mathbf{H}_m(t)\in\mathbb{R}^{9M\times N_{r}}$ be $\delta_m(t)\mathbf{B}_m\mathbf{H}_m(t)=\mathbf{T}_{1,m}^T(t)\Xi_{1,m}(t)\mathbf{S}_{1,m}(t)$, where $\mathbf{T}_{1,m}(t)\in\mathbb{R}^{9M\times 9M}$ and $\mathbf{S}_{1,m}(t)\in\mathbb{R}^{N_{r}\times N_{r}}$ are unitary matrices, and $\Xi_{1,m}(t)\in\mathbb{R}^{9M\times N_{r}}$ is a rectangular matrix containing the singular values $\left\{\sigma_{m,1}(t),...\sigma_{m,\text{Rank}(\delta_m(t)\mathbf{B}_m\mathbf{H}_m(t))}(t)\right\}$. Define $\Sigma_m(t)=\mathbf{T}_{1,m}^T(t)\Sigma(t)\mathbf{T}_{1,m}(t)$ and $\zeta_m(t)=\mathbf{T}_{1,m}^T(t)$  $\Diag((\sigma_{m,1}(t))^{-2}, (\sigma_{m,2}(t))^{-2},..., (\sigma_{m,\text{Rank}(\delta_m(t)\mathbf{B}_m\mathbf{H}_m(t))}(t)$ $)^{-2}, 0,...,0)\mathbf{T}_{1,m}(t)\in\mathbb{R}^{9M\times 9M}$. The dynamic communication and control solutions to Problem 1 are then encapsulated in the following Theorem 2.

\begin{theorem}
\emph{(Semantic Solution)}
The communication and tracking control solution to Problem 1 is given as follows.
\begin{itemize}
    \item \textbf{Inactive Mode:} If 
    \begin{equation}\label{eq10}
       P_{on}\geq \Tr(\Pi\Sigma_m(t)(M\Sigma(t)+\gamma\zeta_m(t))^{\mathcal{y}}\Sigma_m^T(t)\Pi^T),
    \end{equation}
  then   $\delta_m^*(t)=0, \mathbf{K}_m^*(t)=\mathbf{0}_{9M\times N_{r}}$.
    
\item  \textbf{Operative Mode:}  If
\begin{equation}\label{eq11}
     P_{on}< \Tr(\Pi\Sigma_m(t)(M\Sigma(t)+\gamma\zeta_m(t))^{\mathcal{y}}\Sigma_m^T(t)\Pi^T),
\end{equation}
then $\delta_i^*(t)=1$ and
    \begin{equation}\label{eq12}
    \begin{split}
       &\!\!\!\!\!\!\!\!\!\!\!\mathbf{K}_m^*(t)=\mathbf{V}_m^T(t)\Xi_m^{\mathcal{y}}(t)\mathbf{U}_m(t)\Pi\Sigma(t)(M\Sigma(t)+\gamma\zeta_m(t))^{\mathcal{y}}.
    \end{split}
    \end{equation}
\end{itemize}
\end{theorem}

\begin{IEEEproof}
See Appendix B.
\end{IEEEproof}

The optimal communication solution $\delta_m^*(t)$ and the tracking control solution $\mathbf{K}_m^*(t)$  in Theorem 2 exhibit a decentralized semantic structure. Specifically, a significant tracking error cost (when $\|\Sigma(t)\|$ is large) necessitates the transmission of the control signal $\mathbf{u}_m(t)=-\mathbf{K}_m^*(t)(\mathbf{x}(t)-\mathbf{r}(t))$ with a correspondingly large $\mathbf{K}_m^*(t)$ from the leader UAV to the $m$-th follower UAV. Additionally, a favorable channel condition (when $\|\mathbf{H}_m(t)\|$ is large) lead to the transmission of $\mathbf{u}_m(t)=-\mathbf{K}_m^*(t)(\mathbf{x}(t)-\mathbf{r}(t))$ with a smaller $\mathbf{K}_m^*(t)$ from the leader UAV to the $m$-th follower UAV.

\subsection{Algorithm Design}
We summarize the decentralized semantic communication and tracking control algorithm in the following  Algorithm \ref{alg1}.

\begin{algorithm}
\small
\caption{Semantic Communication and Tracking Control for the UAV Swarm over MIMO Fading Channels}
\label{alg1}

\textbf{Step 1} (\emph{Information Broadcasting}): At each $t$-th timeslot, each $m$-th follower UAV ($1\leq m\leq M$)  broadcasts the plant state $\mathbf{x}_m(t)\in\mathbb{R}^{9\times 1}$ to the leader UAV. The leader UAV broadcasts the common pilot $\mathbf{T}\in\mathbb{R}^{N_{t}\times N_t}$ to all follower UAVs.

\textbf{Step 2} (\emph{Channel Estimation and Channel Feedback}): Each $m$-th follower UAV  obtains the channel state  $\hat{\mathbf{H}}_m(t)$ by performing the channel estimation using  the received pilot signal $\mathbf{Y}_m^p(t)=\mathbf{H}_m(t)\mathbf{T}+\mathbf{V}^p(t)\in\mathbb{R}^{N_r\times N_t}$  ($1\leq m\leq M$) from the leader UAV, where  $\mathbf{V}^p(t)\sim\mathcal{N}(0,\mathbf{1}_{N_{r}})$ is the additive Gaussian noise. The local CSI $\hat{\mathbf{H}}_m(t)$ is then fed back to the leader UAV.

\textbf{Step 3} (\emph{Generation of Semantic Solution}): The leader UAV generates the communication policy $\left\{\delta_1^*(t),...,\delta_M^*(t)\right\}$ for $M$ follower UAVs according to Theorem 2. If $\delta_m^*(t)=1$, the leader UAV further generates the control signal $\mathbf{u}_m^*(t)=-\mathbf{K}_m^*(t)\mathbf{x}(t)$, and transmit the signal to the $m$-th follower UAV through the MIMO channels.

\textbf{Step 3} (\emph{Semantic Tracking Control}): Each $m$-th follower UAV receives the control signal $\hat{\mathbf{u}}_m(t)$ and adjusts its state  according to \eqref{eq1}. Let $t=t+1$ and proceed to \textbf{Step 1}.

\end{algorithm}

\subsection{Sufficient Condition for Tracking Stability}
The sufficient condition for tracking stability can be obtained by analyzing the criteria for
the negative Lyapunov drift in \eqref{eq8} under the proposed scheme. This is formally summarized in the following Theorem.

\begin{theorem}
\emph{(Sufficient Condition for Tracking Stability)}
Let the SVD of $\delta_m(t)\mathbf{B}_m\mathbf{H}_m(t)\mathbf{H}_m^T(t)\mathbf{B}_m^T\in\mathbb{R}^{9M\times 9M}$ be $\delta_m(t)\mathbf{B}_m\mathbf{H}_m(t)\mathbf{H}_m^T(t)\mathbf{B}_m^T=\mathbf{T}_{2,m}(t)\Xi_{2,m}(t)\mathbf{S}_{2,m}(t)$, where $\mathbf{T}_{2,m}(t)\in\mathbb{R}^{9M\times 9M}$ and $\mathbf{T}_{2,m}(t)\in\mathbb{R}^{N_r\times N_r}$ are unitary matrices. Let $\mathbf{M}_m(t)\in\mathbb{R}^{9M\times 9M}$
be the mask matrix for $\Xi_{2,m}(t)$ satisfying $\mathbf{M}_m(t)\odot \Xi_{2,m}(t)= \Xi_{2,m}(t)$.
Then, if 
\begin{equation}\label{eq13}
  \|\mathbf{1}_{9M\times 9M}-\frac{1}{M}\sum_{m=1}^M\mathbf{M}_m(t)\|<\frac{1}{\alpha},
\end{equation}
the UAV swarm is tracking stable, i.e., $\limsup_{T\rightarrow\infty}\frac{1}{T}\sum_{t=1}^T\mathbb{E}[\|\mathbf{x}(t)-\mathbf{r}(t)\|^2]<\infty$.
\end{theorem}

\begin{IEEEproof}
See Appendix C.
\end{IEEEproof}

\section{Numerical Results}
In this section, we evaluate the performance benefits of the proposed algorithm. Specifically, we compare our scheme against the following baselines:

\begin{itemize}
\item \textbf{Baseline 1} \emph{(Periodic Communication and PID-based  Control):} The tracking control signal $\mathbf{u}_m(t)$ for each timeslot is given by $\mathbf{u}_m(t)=\mathbf{K}_{m,p}(\mathbf{x}(t)-\mathbf{r}(t))+\mathbf{K}_{m,i}\sum_{i=0}^t(\mathbf{x}(i)-\mathbf{r}(i))+\mathbf{K}_{d,i}(\mathbf{x}(t)-\mathbf{r}(t)-(\mathbf{x}(t-1)+\mathbf{r}(t-1)))$, where the proportional gain $\mathbf{K}_{m,p}(t)\in\mathbb{R}^{N_t\times 9M}$, the integration gain $\mathbf{K}_{m,i}(t)\in\mathbb{R}^{N_t\times 9M}$ and the differentiation gain $\mathbf{K}_{m,d}(t)\in\mathbb{R}^{N_t\times 9M}$ are tuned offline using pole placement. The communication of $\mathbf{u}_m(t)$ from the leader UAV to the $m$-th follower UAV occurs when $t\mod T=0$, with the triggering period $T$ satisfying $1\leq T\leq M$.

\item \textbf{Baseline 2} \emph{(State-Dependent Communication and PID-based  Control):} The tracking control signal $\mathbf{u}_m(t)$ mirrors that of Baseline 1. Communication for $\mathbf{u}_m(t)$ is triggered only if $\|\mathbf{x}(t)-\mathbf{r}(t)-\mathbf{x}^l(t)+\mathbf{r}^l(t)\|^2\leq \sigma \|\mathbf{x}(t)-\mathbf{r}(t)\|^2$, where $\mathbf{x}^l(t)\in\mathbb{R}^{9M\times 1}$ and $\mathbf{r}^l(t)\in\mathbb{R}^{9M\times 1}$ represent the plant state and target at the last triggering timeslot, respectively.
   
\item  \textbf{Baseline 3} \emph{(State-Dependent Communication and GARE-based  Control):} The tracking control solution $\mathbf{u}_m(t)=\mathbf{K}_{m,g}(t)(\mathbf{x}(t)-\mathbf{r}(t))$, where the static tracking control gain $\mathbf{K}_{m,g}\in\mathbb{R}^{N_t\times 9M}$ is obtained via solving the static game algebraic Riccati equation  (GARE) for the UAV swarm over the static channels, i.e., $\mathbf{H}_m(t)=\mathbf{1}_{N_r\times N_t}$, in an offline manner.
   
\end{itemize}

We consider a typical UAV swarm featuring the internal transition matrix $\mathbf{A}_{mm}\in\mathbb{R}^{9\times 9}$   actuation matrix $\mathbf{B}_m\in\mathbb{R}^{9\times 9}$,  both randomly generated following a Gaussian distribution with zero mean and unit variance.  For any $m\in\left\{1,2,...,M\right\}$, the dynamic relationship between the $n$-th follower UAV and the $m$-th follower UAV is defined by $\mathbf{A}_{m,n}=\mathbf{A}_{m,m}, n=(m\mod M) +1$ and $\mathbf{A}_{m,n}=\mathbf{0}_{9}, n\not=m,n\not=(m\mod M) +1$. The  plant noise $\mathbf{w}_m(t)$ follows a Gaussian distribution $\sim\mathcal{N}(\mathbf{0}_{9},10^{-5}\mathbf{I}_{9})$. The triggering constant $\sigma_m=m, 1\leq m\leq M$, and the triggering period $\mathbf{T}=\lceil\frac{M}{2}\rceil$. The target transition matrix is $\mathbf{G}=\mathbf{I}_{9M}$. Initial states are set with $\mathbf{x}(0)=[1,...,1]^T\in\mathbb{R}^{9\times 1}$ and $\mathbf{r}(0)=[100,...,100]^T\in\mathbb{R}^{9\times 1}$

\subsection{Impact of the Number of UAVs} 
\begin{figure}
    \centering
    \includegraphics[height=3.9cm,width=7.2cm]{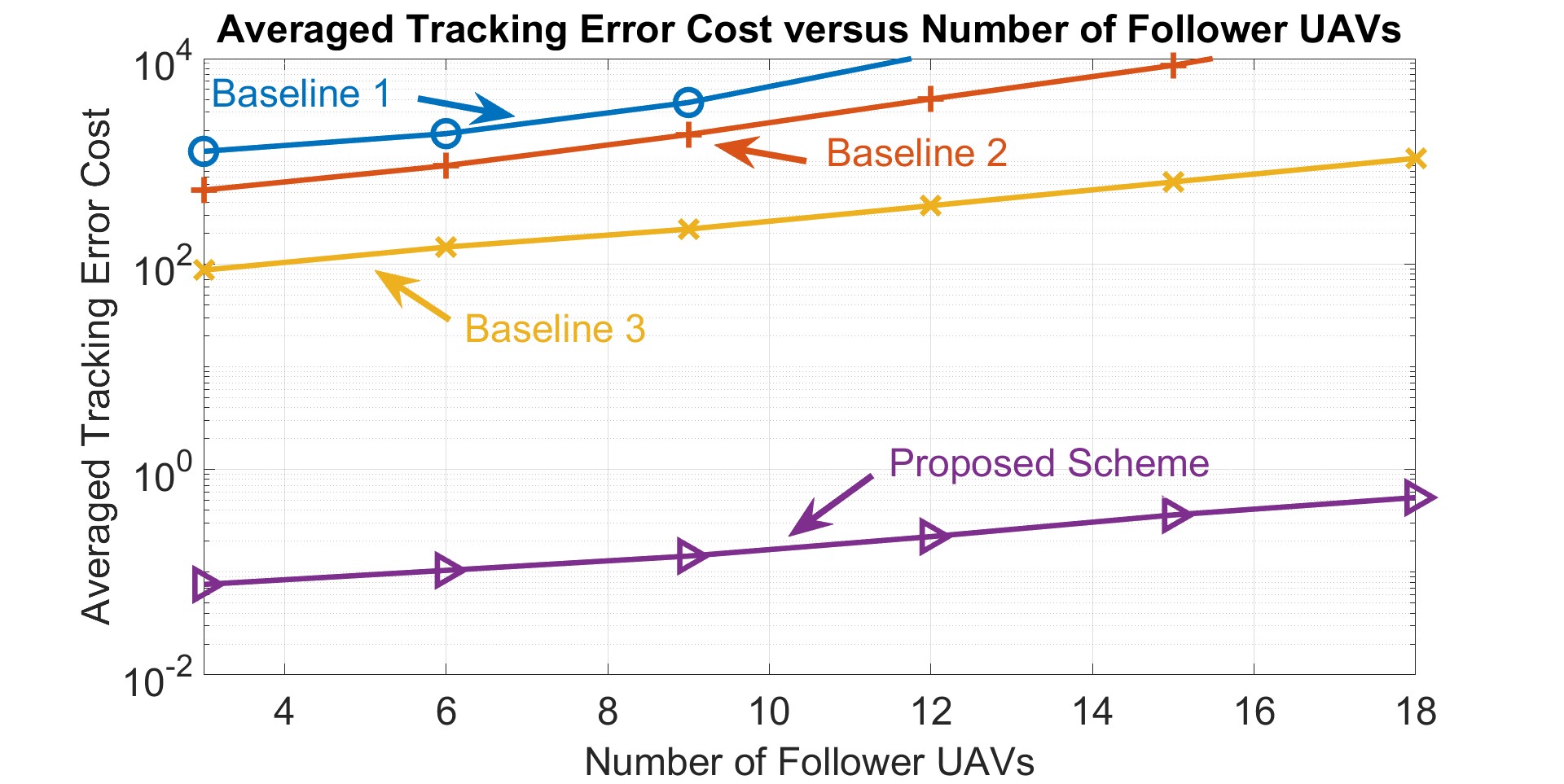}
    \caption{Averaged tracking error cost versus number of follower UAVs where $N_t=12$ and the averaged transmission power is 8 dBW.}
    \label{fig2}
\end{figure}
Fig. \ref{fig2} depicts the average tracking error cost $\mathbb{E}[\|\Sigma(t)\|]$ across $10^4$ timeslots in relation to the number of follower UAVs.  The figure indicates that the average tracking error cost increases for all schemes as the number of follower UAVs grows, reflecting the heightened system instability that accompanies a larger swarm. Notably, our proposed scheme consistently attains a lower average tracking error cost when compared to the baseline schemes, which do not account for real-time CSI in their communication and tracking control algorithm designs.

\subsection{Impact of the Number of Transmission Antennas} 
\begin{figure}
    \centering
    \includegraphics[height=4cm,width=7.2cm]{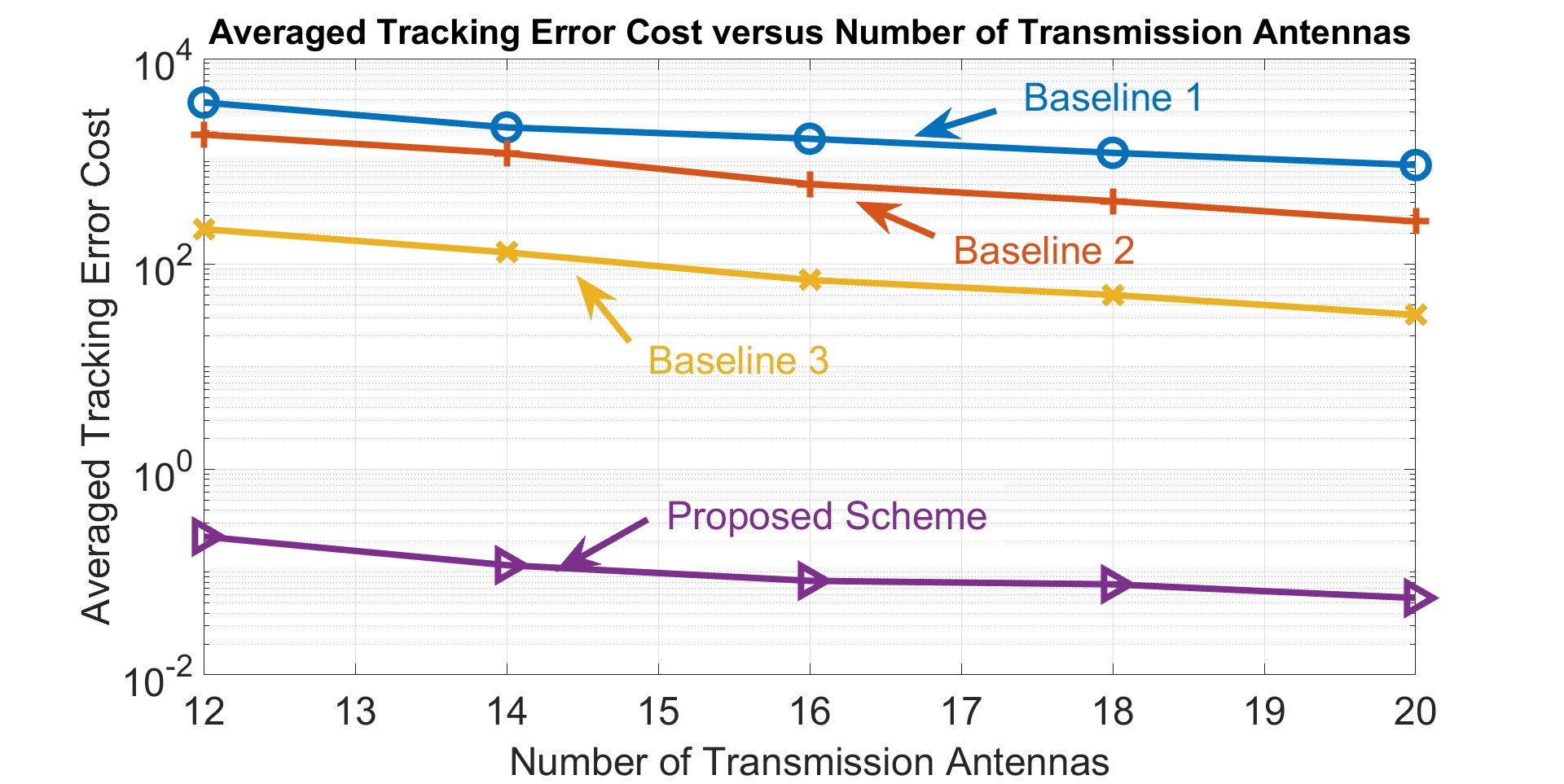}
    \caption{Averaged tracking error cost versus number of transmission antennas where $M=12$ and the averaged transmission power is 8 dBW. }
    \label{fig3}
\end{figure}
Fig. \ref{fig3} presents the average tracking error cost $\mathbb{E}[\|\Sigma(t)\|]$ plotted against the number of transmission antennas $N_t$ across $10^4$ timeslots. The results indicate that an increase in $N_t$ leads to a reduction in tracking error cost for all schemes, as the expanded communication resources enhance the overall system performance. Our proposed scheme excels over the baseline schemes by dynamically adjusting the communication and control strategies in response to real-time plant and channel state realizations.

\subsection{Impact of Averaged Transmission Power at the leader UAV}
\begin{figure}
    \centering
    \includegraphics[height=4cm,width=7.2cm]{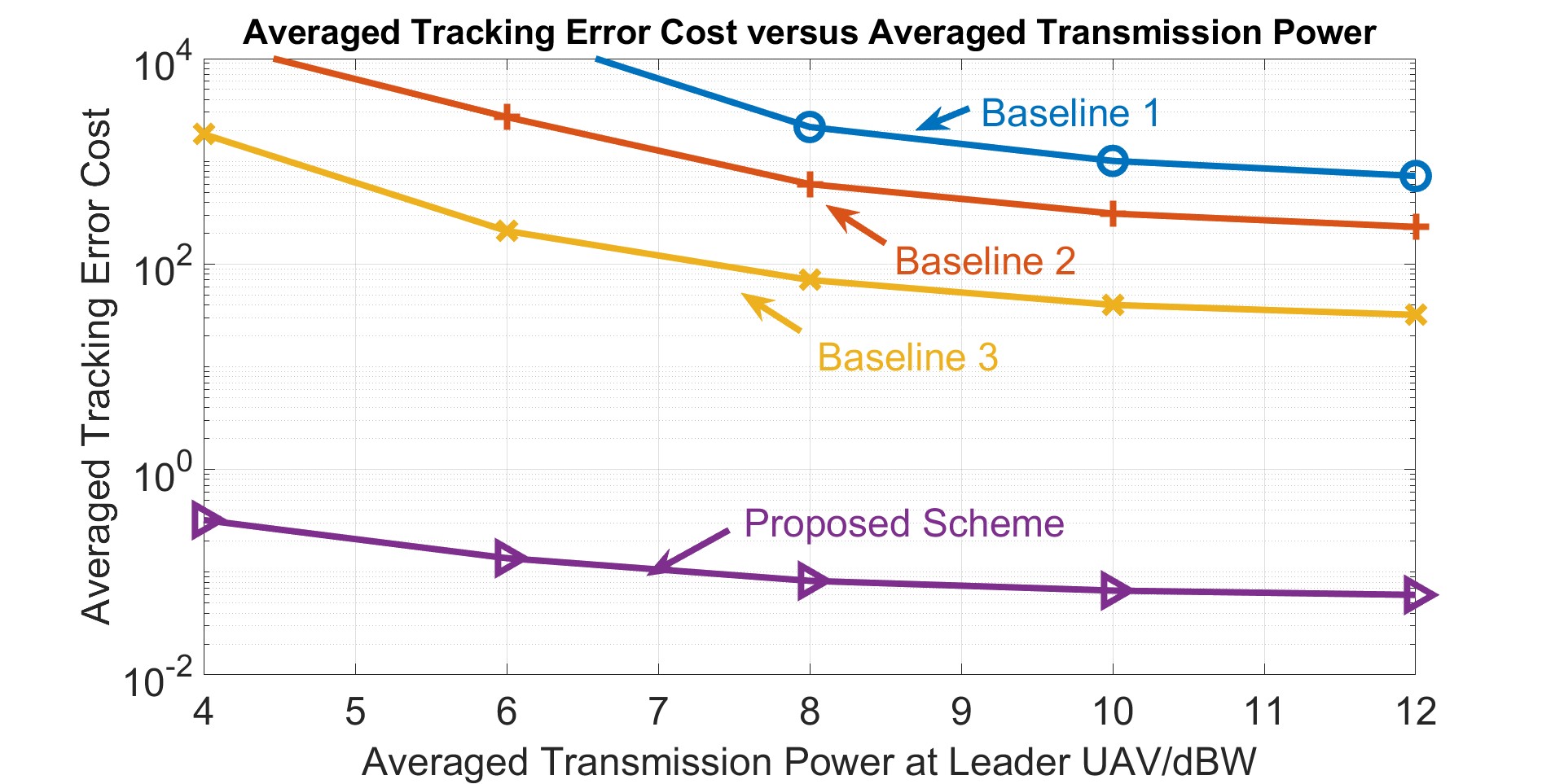}
    \caption{Averaged tracking error cost versus averaged transmission power where $M=N_t=12$.}
    \label{fig4}
\end{figure}
Fig. \ref{fig4} illustrates the average tracking error cost $\mathbb{E}[\|\Sigma(t)\|]$ for the UAV system in relation to the average transmission power at the leader UAV. It is observed that across all schemes, the tracking control performance improves with an increase in transmission power, attributed to the higher received SNRs at the follower UAVs. Notably, our proposed scheme demonstrates a marked advantage in power efficiency, achieving tracking stability compared to the baseline schemes.

\section{Conclusions}
In this work, we delved into the realms of decentralized semantic communication and cooperative tracking control for a UAV swarm over wireless MIMO fading channels. We first modeled the UAV swarm by considering the internal dynamics and communication channels in the swarm. Subsequently, we formulated the communication and tracking control problem for the UAV swarm as a power-aware drift-plus-penalty problem and derived a closed-form solution that maintains a decentralized semantic structure. This structure allows the communication and tracking control policies to adapt to the local CSIs and plant states. Using the Lyapunov drift approach, we derived a closed-form sufficient condition for tracking stability. Numerical results showed that our proposed scheme outperforms surpasses various cutting-edge methods.

\appendix

\subsection{Proof of Theorem 1}
Substitute (2), (3), (4) and (6) into (7), we have:
\begin{equation}
\begin{split}
&\Gamma(\Sigma(t))=\mathbb{E}\left[L(\Sigma(t+1)) -L(\Sigma(t))|\Sigma(t)\right]\\
=& \mathbb{E}\bigg[\Tr\bigg(\widetilde{\mathbf{x}}^T(t)(\widetilde{\mathbf{A}}-\sum_{m=1}^M\widetilde{\mathbf{H}}_m(t)\widetilde{\mathbf{K}}_m(t))^T\mathbf{Q}(\widetilde{\mathbf{A}}-\sum_{m=1}^M\widetilde{\mathbf{H}}_m(t)\\&\widetilde{\mathbf{K}}_m(t))\bigg)-\Tr(\Sigma(t))+\Tr(\mathbf{W})+\sum_{m=1}^M\Tr(\mathbf{B}_m\mathbf{B}_m^T)|\Sigma(t)\bigg]\\
=&\mathbb{E}\bigg[\Tr(\mathbf{W})+\sum_{m=1}^M\Tr(\mathbf{B}_m\mathbf{B}_m^T)+\Tr\bigg(\widetilde{\mathbf{x}}^T(t)\widetilde{\mathbf{A}}^T\mathbf{Q}\widetilde{\mathbf{A}}\widetilde{\mathbf{x}}(t)\bigg)\\-&2\Tr\bigg(\widetilde{\mathbf{x}}^T(t)\widetilde{\mathbf{A}}^T\mathbf{Q}(\sum_{m=1}^M\widetilde{\mathbf{H}}_m(t)\widetilde{\mathbf{K}}_m(t))\mathbf{x}(t)\bigg)-\Tr(\Sigma(t))+\Tr\\
&\bigg( \widetilde{\mathbf{x}}^T(t)(\sum_{m=1}^M\widetilde{\mathbf{H}}_m(t)\widetilde{\mathbf{K}}_m(t))^T\mathbf{Q}(\sum_{m=1}^M\widetilde{\mathbf{H}}_m(t)\widetilde{\mathbf{K}}_m(t))\mathbf{x}(t)\bigg)|\Sigma(t)\bigg],
\end{split}
\end{equation}
where $\widetilde{\mathbf{x}}(t)=[\mathbf{x}^T(t),\mathbf{r}^T(t)]^T\in\mathbb{R}^{18M\times 1}$, $\widetilde{\mathbf{A}}=\Diag(\mathbf{A},\mathbf{G})\in\mathbb{R}^{18M\times 18M}$, $\mathbf{Q}=\begin{bmatrix}
    \mathbf{I}_{9M}&-\mathbf{I}_{9M}\\ -\mathbf{I}_{9M} &\mathbf{I}_{9M}
\end{bmatrix}\in\mathbb{S}^{18M}$, $\widetilde{\mathbf{H}}_m(t)=\begin{bmatrix}
    \delta_m(t)\mathbf{B}_m\mathbf{H}_m(t)&\mathbf{0}_{9M\times N_t}\\\mathbf{0}_{9M\times N_t}& \mathbf{0}_{9M\times N_t}
\end{bmatrix}\in\mathbb{R}^{18M\times 2N_t}$, and $\widetilde{\mathbf{K}}_m(t)=\begin{bmatrix}
    \mathbf{K}_m(t)&-\mathbf{K}_m(t)\\ \mathbf{0}_{N_t\times 9M}&\mathbf{0}_{N_t\times 9M}
\end{bmatrix}\in\mathbb{R}^{2N_t\times 18M}$.

Observing the fact that:
\begin{equation}
    \begin{split}  \Tr(\widetilde{\mathbf{x}}^T(t)\widetilde{\mathbf{A}}^T\mathbf{Q}\widetilde{\mathbf{A}}\widetilde{\mathbf{x}}(t))\leq \alpha\Tr(\Sigma(t)),  
    \end{split}
\end{equation}
and
\begin{equation}
\begin{split}
    &\Tr(\widetilde{\mathbf{x}}^T(t)\widetilde{\mathbf{A}}^T\mathbf{Q}(\sum_{m=1}^M\widetilde{\mathbf{H}}_m(t)\widetilde{\mathbf{K}}_m(t))\mathbf{x}(t))\\
\geq & \sum_{m=1}^M\Tr(\delta_m(t)\mathbf{B}_m\mathbf{H}_m(t)\mathbf{K}_m(t)\Sigma(t)\Pi).
\end{split}
\end{equation}
This  results in that the Lyapunov drift (14) can be characterized by
\begin{equation}
\begin{split}
&\Gamma(\Sigma(t)) \leq   \Tr(\mathbf{W})-\mathbb{E}\bigg[\Tr(\Sigma(t))+\alpha\Tr(\Sigma(t))\\
&-2\sum_{m=1}^M\Tr\bigg(\delta_m(t)\mathbf{B}_m\mathbf{H}_m(t)\mathbf{K}_m(t)\Sigma(t)\Pi\bigg)+\sum_{m=1}^M\Tr(\mathbf{B}_m\\
&\mathbf{B}_m^T)+\Tr\bigg(\delta_m(t)\Sigma(t)\big(\sum_{m=1}^M\delta_m(t)\mathbf{B}_m\mathbf{H}_m(t)\mathbf{K}_m(t)\big)^T\\&\big(\sum_{m=1}^M\mathbf{B}_m\mathbf{H}_m(t)\mathbf{K}_m(t)\big) \bigg)|\Sigma(t)\bigg].
\end{split}
\end{equation}
Further note that for any given two same dimensional matrices $\mathbf{M}_1$ and $\mathbf{M}_2$, the matrix $(\mathbf{M}_1-\mathbf{M}_2)(\mathbf{M}_1-\mathbf{M}_2)^T$ is a positive semi-definite matrix. As a result,  $  \mathbf{M}_1\mathbf{M}_2^T+\mathbf{M}_2\mathbf{M}_1^T\leq \mathbf{M}_1\mathbf{M}_1^T+\mathbf{M}_2\mathbf{M}_2^T$, and (17) gives:
\begin{equation}
\begin{split}
    &\!\Gamma(\Sigma(t))\leq \Tr(\mathbf{W})-\Tr(\Sigma(t))+\alpha\Tr(\Sigma(t))-\mathbb{E}\bigg[2\sum_{m=1}^M\Tr(\\&\delta_m(t)\mathbf{B}_m\mathbf{H}_m(t)\mathbf{K}_m(t)\Sigma(t)\Pi)+\sum_{m=1}^M M\delta_m(t)\Tr(\Sigma(t)(\mathbf{B}_m\\&\mathbf{H}_m(t)\mathbf{K}_m(t))^T(\mathbf{B}_m\mathbf{H}_m(t)\mathbf{K}_m(t)))|\Sigma(t)\bigg]+\sum_{m=1}^M\Tr(\mathbf{B}_m\mathbf{B}_m^T).
\end{split}
\end{equation}
This concludes the proof for Theorem 1.

\subsection{Proof of Theorem 2}

First, we evaluate the optimal tracking gain $\mathbf{K}_m^*(t)$ for the $m$-th follower UAV when the communication between the leader UAV and the $m$-th follower UAV is operative, i.e., $\delta_m(t)=1$.
Let $\widehat{\mathbf{K}}_m(t)=\delta_m(t)(t)\mathbf{B}_m\mathbf{H}_m(t)\mathbf{K}_m(t)\in\mathbb{R}^{9M\times 9M}$. The objective function in optimization Problem 1 can be expressed as follows:
\begin{equation}
\begin{split} &f(\widehat{\mathbf{K}}_m(t))=-2\Tr(\Pi\widehat{\mathbf{K}}_m(t)\Sigma(t))+M\Tr(\widehat{\mathbf{K}}_m(t)\Sigma(t)\\&\widehat{\mathbf{K}}_m^T(t))+\delta_m(t)(P_{on}+\gamma\Tr(\widehat{\mathbf{K}}_m(t)\zeta_m(t)\widehat{\mathbf{K}}_m^T(t))).
\end{split}
\end{equation}
The minimum value of $f(\widehat{\mathbf{K}}_m(t))$ when $\delta_m(t)=1$ can be obtained via setting the derivative of $f(\widehat{\mathbf{K}}_m(t))$ w.r.t. the objective variable $\widehat{\mathbf{K}}_m(t)$ to be zero while keeping $\delta_m(t)=1$, given by:
\begin{equation}
    \begin{split}
        &\frac{\partial f(\widehat{\mathbf{K}}_m(t))}{\partial \widehat{\mathbf{K}}_m(t)}|_{\delta_m(t)=1}\\
        =&-2\Pi\Sigma(t)+2M\widehat{\mathbf{K}}_m(t)[\Sigma(t)+\gamma\zeta_m(t)]=0.
    \end{split}
\end{equation}
It gives that:
\begin{equation}
    \begin{split}
        \widehat{\mathbf{K}}_m^*(t)|_{\delta_m(t)=1}=\Pi\Sigma(t)(M\Sigma(t)+\gamma\zeta_m(t))^{\mathcal{y}}
    \end{split}
\end{equation}
and 
\begin{equation}
    \begin{split}
        \mathbf{K}_m^*(t)|_{\delta_m(t)=1}=\mathbf{V}_m^T(t)\Xi_m^\mathcal{y}(t)\mathbf{U}_m(t)\Pi\Sigma(t)(M\Sigma(t)+\gamma\zeta_m(t))^\mathcal{y}.
    \end{split}
\end{equation}
 In other words, the optimal control gain $  \mathbf{K}_m^*(t)$ for the $m$-th UAV is given by $ {\widehat{\mathbf{K}}_m^*(t)}|_{\delta_m(t)=1}$
if the communication between the leader UAV and the $m$-th follower UAV is operative.

Second, we determine the communication condition for the $m$-th follower UAV. Specifically, we substitute the optimal control gain $\mathbf{K}^*_m(t)$ when the communication from the leader UAV to the $m$-th follower UAV is operative into the objective function (19). It gives that
\begin{equation}
    \begin{split}
        &\min f(\widehat{\mathbf{K}}_m(t))|_{\delta_m(t)=1}\\&=f(\widehat{\mathbf{K}}_m^*(t))|_{\delta_m(t)=1}\\&=P_{on}-\Tr(\Pi\Sigma(t)(M\Sigma(t)+\gamma\zeta_m(t))^\mathcal{y}\Sigma(t)\Pi).
    \end{split}
\end{equation}
 Further note that
 \begin{equation}
     \begin{split}
         \min_{\widehat{\mathbf{K}}_m(t)} f(\widehat{\mathbf{K}}_m(t))|_{\delta_m(t)=0}=f(\mathbf{0}_{9M})=0
     \end{split}
 \end{equation}
 Thus, the communication between the leader UAV and the $m$-th follower UAV is operative if and only if:
 \begin{equation}
     \begin{split}
         \min_{\widehat{\mathbf{K}}_m(t)} f(\widehat{\mathbf{K}}_m(t))|_{\delta_m(t)=1}<\min_{\widehat{\mathbf{K}}_m(t)}f(\widehat{\mathbf{K}}_m(t))|_{\delta_m(t)=0}
     \end{split}
 \end{equation}
This concludes the proof for Theorem 2.

\subsection{Proof of Theorem 3}

We substitute the optimal communication policy $\delta_m^*(t)$ and the optimal tracking control policy $\mathbf{K}_m^*(t)$ in Theorem 2 to the Lyapunov drift (18), it gives
\begin{equation}
\begin{split}
    &\quad  \Gamma(\Sigma(t);\left\{\delta_m^*(t)\right\},\left\{\mathbf{K}_m^*(t)\right\})\\
    & + \sum_{m=1}^M\mathbb{E}\bigg[\delta_m^*(t)(P_{on}+\Tr(\mathbf{K}_m^*(t)(\mathbf{K}_m^*(t))^T)|\Sigma(t)\bigg]\\ 
\leq &\Tr(\mathbf{W})+M\sum_{k=1}^K\Tr(\mathbf{B}_k\mathbf{B}_k^T)+\alpha\Tr(\Sigma(t))\\
&-\Tr(\Sigma(t))+\mathbb{E}\bigg[\sum_{m=1}^M\min_{\widehat{\mathbf{K}}_m(t)}f(\widehat{\mathbf{K}}_m(t))|_{\delta_m(t)=1}|\Sigma(t)\bigg].
\end{split}
\end{equation}

We construct a bridging  tracking control gain $\Bar{\mathbf{K}}_m(t)\in\mathbb{R}^{9M\times 9M}$ for the $m$-th follower UAV, given by:
\begin{equation}
    \begin{split}
        \Bar{\mathbf{K}}_m(t)=&\Pi\mathbf{M}(t)\Sigma(t)\mathbf{M}(t)(M\mathbf{M}(t)\Sigma(t)\mathbf{M}(t)\\
        &+\gamma\zeta_m(t))^\mathcal{y}\mathbf{M}(t).
    \end{split}
\end{equation}
Substituting $\Bar{\mathbf{K}}_m(t)$ into (19), the objective function $f(\Bar{\mathbf{K}}_m(t))|_{\delta_m(t)=1}$ is given by
\begin{equation}
\small
    \begin{split}
        &f(\Bar{\mathbf{K}}_m(t))|_{\delta_m(t)=1}=P_{on}-\frac{\|\Pi\|^2}{M}\Tr(\mathbf{M}_m(t)\Sigma(t)\mathbf{M}_m(t))\\
        &+\frac{\|\mathbf{A}\|^2}{M}\Tr( (\mathbf{M}_m(t)\Sigma(t)\mathbf{M}_m(t))^\mathcal{y}\\
        &+\frac{M}{\gamma\Tr( (\mathbf{B}_m\mathbf{H}_m(t))(\mathbf{B}_m\mathbf{H}_m(t))^T)^\mathcal{y}}\mathbf{1}_{9M})^{-1}\\
        \leq &P_{on}-\frac{\|\Pi\|^2}{M}\Tr(\mathbf{M}_m(t)\Sigma(t)\mathbf{M}_m(t))\\
        &+\gamma\Tr( (\mathbf{B}_m\mathbf{H}_m(t))(\mathbf{B}_m\mathbf{H}_m(t))^T)^\mathcal{y}.
    \end{split}
\end{equation}
Since $\widehat{\mathbf{K}}^*_m(t)$ is the optimizer for the function $f(\cdot)|_{\delta_m(t)=1}$, i.e., $f(\widehat{\mathbf{K}}_m^*(t))=\min_{\widehat{K}_m^*(t)}f(\widehat{\mathbf{K}}_m^*(t))|_{\delta_m(t)=1}$, the Inequalities (27) and (28) give
\begin{equation}
\begin{split}
    &\Gamma(\Sigma(t);\left\{\delta_m^*(t)\right\},\left\{\mathbf{K}_m^*(t)\right\})\\
    + & \sum_{m=1}^M\mathbb{E}[\delta_m^*(t)(P_{on}+\mathbf{K}_m^*(t)(\mathbf{K}_m^*(t))^T)|\Sigma(t)]\\
    \leq& \Tr(\mathbf{W})+\sum_{m=1}^M\alpha\Tr(\mathbf{B}_m\mathbf{B}_m^T)+MP_{on}\\& +\sum_{m=1}^M\gamma\mathbb{E}\bigg[\Tr\bigg((\mathbf{B}_m\mathbf{H}_m(t)\mathbf{H}_m^T(t)\mathbf{B}_m^T)\bigg)^\mathcal{y}\bigg]\\
    &+\bigg(\alpha\|\mathbf{1}_{9M}-\frac{1}{M}\sum_{m=1}^M\mathbf{M}(t)\|-1)\Tr(\Sigma(t)\bigg).
\end{split}
\end{equation}
Thus, the sufficient condition for ensuring tracking stability is derived from the  condition that the drift in equation (29) must be negative, as detailed below:
\begin{equation}
    \alpha\|\mathbf{1}_{9M}-\frac{1}{M}\sum_{m=1}^M\mathbf{M}(t)\|-1<0.
\end{equation}
This concludes the proof for Theorem 3.

\bibliographystyle{IEEEtran}
\bibliography{IEEEabrv,Bibliography}

\vfill


\end{document}